\pdfoutput=1

\documentclass{elsart}

\usepackage{graphics}
\usepackage{graphicx}
\usepackage{amssymb}
\usepackage{amsmath}
\usepackage{theorem}
\usepackage{color}

\def \d {\delta}
\def \ep {\varepsilon}

\def \vr {\boldsymbol{r}}
\def \vx {\boldsymbol{x}}
\def \vu {\boldsymbol{u}}
\def \ep {\varepsilon}
\def \pd {\partial }
\def \vnabla {\boldsymbol{\nabla}}

\def \D {\mathcal{D}}
\def \calS {\mathcal{S}}

\def \G {\Gamma}
\def \a {\alpha}
\def \b {\beta}
\def \g {\gamma}
\def \e {\epsilon}

\def \o {\Omega}
\def \go {\bar{g}}
\def \da {\dA}
\def \R {\mathbb{R}}
\def \D {\mathcal{D}}
\def \L {\Lambda}
\def \conab { \bar{\nabla}}
\def \GO {\bar{\Gamma}}
\def \G {\Gamma}
\def \A {\mathcal{A}}
\def \hatn {\boldsymbol{\hat{N}}}
\def \vR {\boldsymbol{R}}
\def \V {\boldsymbol{V}}
\def \e {\mathbf{e}}

\newcommand{\dA}{\sqrt{|\go|}}

\newcommand{\deriv}[2]{\frac{d #1}{d #2}}

\newcommand{\brk}[1]{\left( #1 \right)}

\newcommand{\mymat}[1]{\begin{pmatrix} #1 \end{pmatrix}}
\newcommand{\putfig}[2]{\begin{center} \includegraphics[height=#1in]{#2}\end{center}}

\numberwithin{equation}{section} \numberwithin{figure}{section}

\begin{document}

\begin{frontmatter}


\title{Elastic theory of unconstrained non-Euclidean plates}


\author[phys]{E. Efrati,}
\author[phys]{E. Sharon,}
\author[math]{R. Kupferman}
\address[phys]{The Racah Institute of Physics, The Hebrew University, Jerusalem 91904, Israel}
\address[math]{Institute of Mathematics, The Hebrew University, Jerusalem 91904, Israel}

\begin{abstract}
Non-Euclidean plates are a subset of the class of elastic bodies having no stress-free configuration. Such
bodies exhibit residual stress when relaxed from all external constraints, and may assume complicated equilibrium
shapes even in the absence of external forces. In this work we present a mathematical framework for such bodies
in terms of a covariant theory of linear elasticity, valid for large displacements. We propose the concept of
non-Euclidean plates to approximate many naturally formed thin elastic structures. We derive a thin plate
theory, which is a generalization of existing linear plate theories, valid for large displacements but small
strains, and arbitrary intrinsic geometry.
We study a particular example of a hemispherical plate. We show the occurrence of a spontaneous buckling transition from a stretching dominated configuration to bending dominated configurations, under variation of the plate thickness.
\end{abstract}


\end{frontmatter}

\section{Introduction}
\label{intro}

Elasticity theory, in its most fundamental formulations, describes the statics and dynamics of three-dimensional
(3D) elastic bodies. Such ``fundamental'' models are extremely complex, due to both  high dimensionality and
nonlinearity. This intrinsic complexity has motivated over the years the development of simplified, or reduced
models of elasticity. In particular, models of lower spatial dimension have been developed to describe the
mechanics of slender bodies, such as columns, shells and plates. These models are based on various
approximations, such as lateral inextensibility, small deflections and small deformations. In particular, the
Kirchhoff-Love assumptions \cite{love} allow the derivation of reduced two-dimensional (2D) theories of plates.
The F\"{o}ppl-Von K\'{a}rm\'{a}n (FVK) plate equations are one of the successful reduced descriptions of plates
mechanics. It expresses the elastic energy of a deformed elastic plate as a sum of stretching and bending
energies of a 2D surface.  The stretching energy, which accounts for in-plane deformations, is linear in the plate
thickness, $h$. The bending energy, which depends on the curvature of the deformed plate, is cubic in $h$.
Other reduced 2D theories
usually bear the same structure, i.e., their energy
is given by the sum of a stretching term and a bending term \cite{Koi66}.
The validity of the dimensional reduction from 3D to 2D models, based on the Kirchhoff-Love assumptions,
has been the subject of many scientific disputes \cite{Koi70}.
Recently, the FVK theory has been derived from a 3D elastic theory by means of an asymptotic expansion \cite{Cia97}.
The stretching and bending terms in the FVK theory have also been derived
as two different vanishing thickness $\G$-limits of the 3D elastic energy \cite{FJM06}.

2D elastic theories
distinguish between two types of thin bodies: plates and shells.
Plates are elastic bodies that bear no structural variation across their thin dimension, and
possess a planar rest configuration.
Shells are elastic bodies that bear structural variations across their thin dimension, and as a result,
possess a non-planar rest configuration.
In both cases the postulated
existence of a stress-free, rest configuration is of paramount importance.

Recent technological developments have extended the range of mechanical structures that can be engineered and
constructed. Plates of nanometer scale thickness can be manufactured \cite{HJJCEM07}, responsive nano-structures
 are being developed \cite{ERMVMG05,HC07}, and the use of shape memory materials that lead to large shape
transformations has been extended \cite{materials}. In addition, the application of mechanics to biological
systems, such as in the study of plant mechanics and motility \cite{FSDM05} and the study of mechanically
induced cell differentiation \cite{PCCCCL04}, is a rapidly developing field.
Such developments have
renewed the interest in elasticity. Several recent theoretical works have focused on the onset of various
mechanical instabilities and the scaling of the generated patterns \cite{HJJCEM07,CM03}, and other thoroughly
analyzed the assumptions underlying some of the dimensionally reduced models \cite{FJM06}.

The modeling of growing elastic bodies is an area  in which current theories of elasticity
face difficulties.
Growing tissues, such as leaves, exhibit very complex configurations even in
the absence of external forces \cite{SMS04}.
Although leaves (and many other growing tissues) are relatively
thin (compared to their lateral dimensions), there are no reduced 2D elastic theories that model their shaping
mechanisms. Another class of systems for which current theories do not apply are elastic bodies undergoing irreversible plastic deformations.
The main difficulty in applying elasticity theory to growing bodies,
or elastic bodies having undergone plastic deformations,
is their \emph{lack of a stress-free configuration}.
Specifically, in most models,
the elastic energy density of a deformed body depends on the
local elastic modulus and the strain tensor. The latter is defined by
the gradient of the mapping between a stress-free configuration and the deformed configuration.
It can be shown, for example, that a general
growth process of an elastic material leads to a body that has no stress-free configuration,
thus exhibiting residual stress in the absence of external loading \cite{GM07}.

To formulate an elastic theory for bodies that do not have stress-free configurations, one needs an alternative definition of the strain tensor.
At present, certain 3D formulations use  the concepts of  \textit{virtual configuration} \cite{BG05,Hog93} and
\textit{intermediate configuration} \cite{Sidoroff02,Sidoroff82} to describe natural growth processes as well as
plastic deformations leading to residual stress. The growth process in these theories is decomposed into a
\emph{growth step}, which  maps a stress-free configuration into a virtual configuration, and an \emph{elastic relaxation step}, which
maps the virtual configuration into an elastic equilibrium configuration that contains residual stress.
These theories use a multiplicative decomposition of the deformation gradient into an elastic and
a plastic part. Other theories decompose the strain tensor additively \cite{GN71}.

In the current work, we focus on the elastic response of the body after its
``rest configuration" has been modified either by growth, or by plastic deformation.
We do not consider  the thermodynamic limitations on plastic deformations (which are
not relevant to naturally growing tissue). We assume that the distorted ``rest configuration" (or
virtual configuration) is a known quantity.
If an elastic body is capable of assuming the virtual configuration, then there exists a stress-free
configuration, which is unique; the solution to the elastic problem is then trivial.
If, however, no elastic body can assume the virtual configuration, then
no stress-free configuration exists, and we face a non-trivial problem which exhibits residual stress.
We term such bodies as ``non-Euclidean" because their internal geometry is not immersible in three-dimensional Euclidean space.

\begin{figure}[h]
\putfig{0.8}{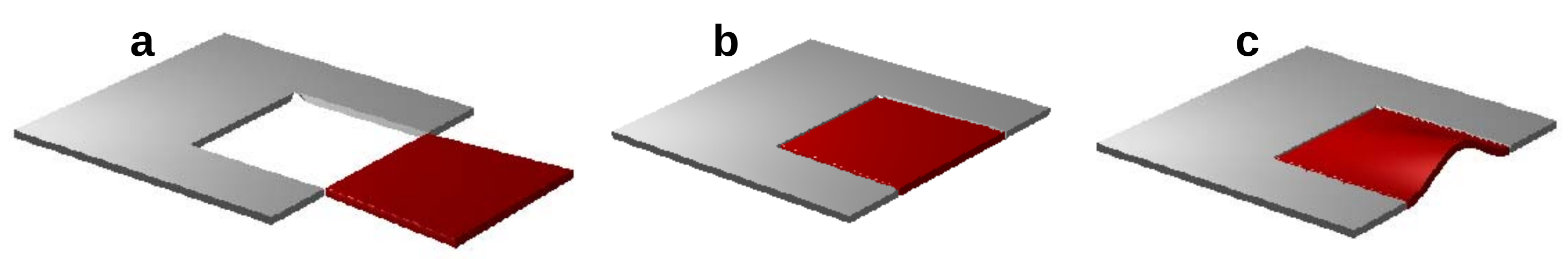}
\caption{
Schematic illustration of an unconstrained plate exhibiting residual stress.
(a) The two elements composing the plate are shown side by side.
(b) As the red trapezoid is too large to fit into the square opening, it is  compressed.
(c) For a plate sufficiently thin, the induced compression exceeds the buckling threshold, and the
trapezoid buckles out of plane . Note that there are many shapes that preserve all lengths along the faces
of the plate, yet they cannot be planar.}
\label{fig:slab}
\end{figure}

We consider now two model examples of elastic structures that belong to the class of systems we have termed non-Euclidean plates, and discuss qualitatively some of their properties. Consider an elastic
square slab of lateral dimensions $2L$, and thickness $h$.
Suppose we cut out from it a square segment of dimension $L$, leaving out a U-shaped structure  (see Figure \ref{fig:slab}$a$). Next, the square is replaced by a trapezoid that has three edges of equal length $L$, and a fourth edge of longer size $L'$. Of course, the trapezoid is too large to fit in the square slot. Suppose, however, that we forcefully insert the trapezoid into the slot, gluing its three sides of length $L$ to the corresponding edges of the U-shape. As a result, the U-shape will slightly open, whereas the trapezoid will experience compression.
This plane-stress configuration is shown schematically in Figure \ref{fig:slab}$b$. If the plates are sufficiently thin,
the trapezoid is unable to sustain the compression and buckles out of plane to form a shape
qualitatively described in Figure \ref{fig:slab}$c$.

We note the following points for this toy problem:

\begin{enumerate}
\item The three dimensional metric that describes the \textit{rest lengths} of the compound body (U-shape plus trapezoid) is continuous.

\item If $x^3$ denotes the vertical coordinate
(say, the distance from the bottom face), then all $x^3=const$ surfaces
are identical. It is this property that causes the body to remain flat (for sufficiently thick samples), and
will later be used to rigorously define non-Euclidean plates.

\item The body exhibits residual stress in the absence of external
constraints: in Figure \ref{fig:slab}$b$ the body is in a state of non-trivial plane-stress, identical for all
$x^3=const$ sections. In the buckled state (Figure \ref{fig:slab}$c$) symmetry is broken. The upper surface is longer then the lower
surface, hence at least one of them must be strained.
It may easily be shown that the \textit{compound} body has no
unstressed configuration.

\item The problem is purely geometric: As both pieces (the confining U and the
trapezoid) are made of the same material,
the stiffness of the material (Young's modulus) has no effect on the equilibrium shape, and we expect to see the same behavior for metals and rubbers (as long as the strains are sufficiently small and the stresses are below the yield stress).

\item The toy problem presented here may easily be solved numerically using commercial software
(In fact, a very similar problem was addressed experimentally and analytically in \cite{MB06}).
The treatment used for solving such problems is
limited to \textit{discrete} geometric incompatibilities: two (or more) regular elastic problems that are
coupled through their boundary conditions are solved simultaneously.
Plastic deformations and non-homogeneous growth processes, however, cannot be
mapped into such discrete geometries.
\end{enumerate}

\begin{figure}[h] \putfig{0.8}{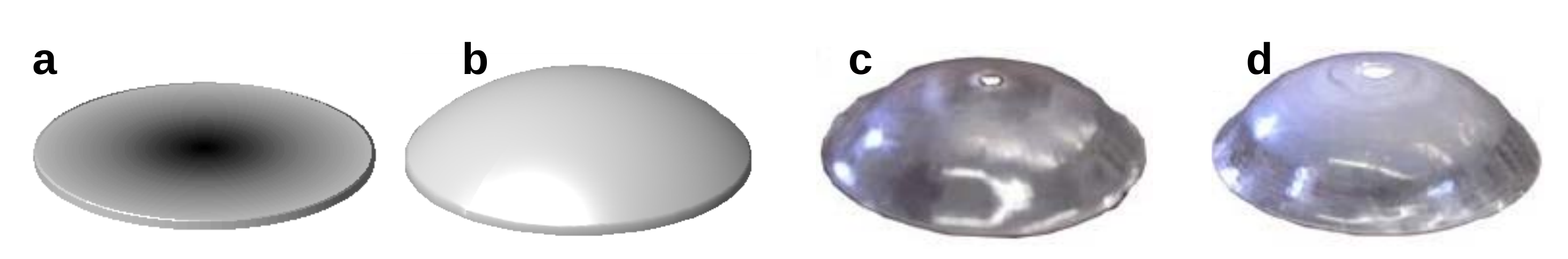}
\caption{
An initially flat disc shrinking differentially. (a) The peripheral areas (light grey) shrink significantly, while the center of the disc (dark grey) shrinks moderately. (b) In order to accommodate the center of the disc within the ``too short" peripheral ring, the plate must buckle out of plane. While the newly prescribed lateral lengths are satisfied on average (over the thickness), the symmetry breaking causes the upper surface to be tensed, while the lower surface is compressed. (c-d) Experimental realization of non-Euclidean plates, using environmentally
responsive gels as described in \cite{KES07}. The differential shrinkage prescribes a two dimensional geometry of  constant positive Gaussian curvature $K = 0.11$cm$^{-2}$. The thicknesses of the gels are $h_c = 0.75$mm and $h_d = 0.6$mm.
}
\label{fig:caps}
\end{figure}

Recent experiments in torn plastic sheets \cite{SRMSS02} and environmentally responsive gel discs \cite{KES07}
have attracted attention to a specific class of non-Euclidean elastic bodies: thin bodies whose shaping mechanism is essentially two-dimensional. Growing leaves display such behavior, as their growth is believed to be nearly homogeneous across their thin dimension, and inhomogeneous in the lateral dimensions.
The gel discs reported in \cite{KES07} mimic a growing thin 3D body shaped by a 2D growth process.
In these experiments initially flat stress-free objects shrink according to a pre-determined chemical
gradient in their composition. The shrinking is homogeneous across the thickness, but inhomogeneous in the
lateral directions (see Figure \ref{fig:caps} for an example). The resulting body shows no structural variation across its thin dimension, yet the lateral equilibrium distances, specified by the differential shrinking, define a 2D non-Euclidean metric tensor. Thus, they cannot be preserved in any flat configuration of the disc. Such bodies may not be considered as plates (due to their non-planar intrinsic geometry), nor as shells (as there are no structural variations across the thin dimension). We name such bodies non-Euclidean plates.

The configurations of non-Euclidean plates in the absence of
external forces are not flat (Figure~\ref{fig:caps}$c$ and \ref{fig:caps}$d$), and may exhibit multi-scale, and fractal-like configurations
\cite{SRMSS02,KES07}. Finite element simulations devised to describe such bodies
\cite{MP06,AB03}, were able to obtain
such multi-scale configurations as energy minima. In both computational and
theoretical works, it was assumed that the elastic energy can be written as a sum of bending and
stretching terms. The bending was measured with respect to a locally flat configuration (as in the FVK plate
model), and the stretching was evaluated with respect to a reference 2D metric tensor. None of these
works, however, was backed up with a theoretical justification for such assumptions.

In the present work we derive a reduced 2D elastic theory for non-Euclidean plates and discuss their
characteristics. The derivation starts from  a model of  a  3D covariant ``incompatible" elasticity, that is, a model for 3D bodies whose intrinsic metric cannot be immersed in a 3D Euclidean space.
We advocate that the common definition of strains with respect to a stress-free configuration
is too restrictive. Instead, strains can be measured with respect to a \emph{reference metric tensor}, which is not necessarily immersible in 3D Euclidean space (incompatibility).
When the strain tensor is defined with respect to a metric tensor, growth
(or any other metric prescription) is naturally decoupled from the elastic relaxation. The second Cauchy-Piola
stress tensor (which is linear in the strain for small strains), may be written explicitly in terms of the
difference between two metric tensors. In such a formulation residual stress appears inevitably as a result of the lack of immersibility.
.

We apply this formulation to thin elastic plates, using the Kirchhoff-Love assumptions. When applied to
ordinary plates, our theory coincides with the Koiter plate theory \cite{Koi66}.
As in the FVK and Koiter theories, the energy of the plate
is a sum of stretching and bending terms. The bending term is cubic in $h$ and quadratic in surface curvatures.
The stretching term is linear in $h$ and depends on the difference between the 2D metric tensor of the
configuration and the reference metric $\go$, (in \cite{MP06} it was termed ``target metric")
The covariant elasticity formulation, together with the bending term measures deviations from a flat configuration, while the stretching term
measures deviations from the 2D reference metric (which may be non-flat).
The resulting model is simple to use, and has an intuitive structure, which clarifies  the underlying physics. We end this paper with an
application of the theory to a simple case of a hemispherical plate.

\section{Theoretical framework: covariant linear elasticity theory}
\label{elastic theory}

In this section we derive the energy functional of a three-dimensional elastic body as a function of its metric
using general curvilinear coordinates. We will show that the energy functional takes the following form,
\[
E(g)=\int_\D w(g)\da dx^1dx^2dx^3 \qquad w=\half A^{ijkl}\ep_{ij}\ep_{kl},
\]
where we use the Einstein summation convention and
\begin{equation}
\label{A} A^{ijkl}=\lambda \go^{ij}\go^{kl}+\mu\left(\go^{ik}\go^{jl}+\go^{il}\go^{jk}\right) \qquad
\ep_{ij}=\half\left(g_{ij}-\go_{ij}\right).
\end{equation}
Here $g_{ij}$ is the metric tensor, $\go_{ij}$ is a symmetric positive-definite tensor, which we term the
\emph{reference metric}, and $\lambda,\mu$ are elasticity (Lam\`{e}) constants; for tensors $|\cdot|$ denotes the determinant.
This energy functional neglects terms that are of order higher than quadratic in $\ep_{ij}$, which is the
deviation of the metric from the reference metric. For bodies which possess a stress-free configuration, $\go$ may
be called the rest metric and must comply with six additional differential constraints (the vanishing of the
Ricci curvature tensor). Precise definitions will be provided in the following subsections. For a thorough
treatment of bodies that have a stress-free configuration, the reader is referred to the recent introductory
book by Ciarlet  \cite{Cia05}, which contains the mathematical background to the subject. A similar treatment,
which we consider as a starting point for our generalization,  can be found in Koiter \cite{Koi66}. We  derive
the energy functional in a slightly different manner, yet  we try as far as possible to use the notations of
\cite{Koi66}, later adopted in \cite{Cia05}.

\subsection{``Incompatible" covariant three-dimensional elasticity}

When a body (a compact  domain $\o\subset\R^3 $) is endowed with a regular set of material curvilinear
coordinates $\vx=(x^1,x^2,x^3)$, it is also endowed with an induced metric tensor. Specifically, if $\vr$
denotes the mapping from the domain of parametrization, $\D\subset\R^3$, into $\o$ (we call $\vr$ the
\emph{configuration} of the body), then the endowed metric is $g_{ij}= \pd_i \vr \cdot \pd _j \vr$. Here and
below we use roman lower-case letters $,i,j,\dots$ for indices $\{1,2,3\}$; the operator $\partial_i$ denotes
the partial derivative with respect to $x^i$. Any deformation of the body (carrying the coordinates along with
every material point) will result in a different metric tensor. A rigidity theorem states that if the induced
metrics of two configurations  $\vr(\vx)\in\o$ and $\tilde{\vr}(\vx)\in\tilde{\o}$ satisfy
$g_{ij}(\vx)=\tilde{g}_{ij}(\vx)$ for every $\vx\in \D$, then the two configurations can only differ by a rigid
motion (a uniform translation and a rigid rotation). Thus, the metric (provided that it is immersible in $\R^3$) uniquely defines the physical configuration of a three-dimensional body.

Our main postulate, which may be viewed as a modification of the hyper-elasticity principle originally
formulated by Truesdell \cite{Tru52}, is:

\begin{verse}
\emph{The elastic energy stored within a deformed elastic body can be written as a volume integral of a local
elastic energy density, which depends only on (i) the local value of the metric tensor, and (ii) local metrial properties  that are  independent of the configuration.}
\end{verse}

The tensors that characterize the material and the body---the \emph{elastic tensors}---contain all the
information about the elastic moduli and the intrinsic geometry of the body. Truesdell's hyper-elasticity
principle is formulated in terms of the strain tensor, which  requires the existence of a stress-free reference
configuration. In contrast, our postulate is formulated in terms of the metric tensor. This obviates the need of
a rest configuration, hence allows for residual stress.

Let $\tilde{w}$ be the energy density per unit volume. The total elastic energy is
\[
E=\int_{\D} \tilde{w} \sqrt{|g|}dx^1dx^2dx^3.
\]
Our postulate states that the function $\tilde{w}$ depends on the metric $g$ and on the coordinates $\vx$
(through the elastic tensors), i.e. $\tilde{w} = \tilde{w}(g,\vx)$. We make the following additional
assumptions:

\begin{enumerate}
\item $\tilde{w}(g,\vx) \ge 0$. \item For every $\vx\in\D$ there exists a unique metric $\go=\go(\vx)$ such that
$\tilde{w}(\go(\vx),\vx) = 0$. We call $\go$ the reference metric.
\end{enumerate}

In the present work we consider the reference metric $\go$ to be a known quantity, whereas the unknown is $g$, the
``actual" metric of the configuration.  It turns out to be more convenient to define the energy density per unit
volume with respect to the volume element induced by the reference metric. We therefore define
$w=\sqrt{|g|/|\go|}\tilde{w}$ as the new energy density. Note that the previous assumptions on $\tilde{w}$ carry
over to $w$, i.e.
\[
w(g,\vx) \ge 0,\qquad w(g,\vx) = 0 \Leftrightarrow g=\go.
\]
If we additionally assume that $w(g,\vx)$ is twice-differentiable with respect to $g$ in the vicinity of $\go$,
then for small deviations of the metric $g$ from the reference metric $\go$ our assumptions imply that
\[
w = \half A^{ijkl} \ep_{ij} \ep_{kl} + O(\ep^4),
\]
where
\[
\ep_{ij} = \half(g_{ij} - \go_{ij})
\]
is the deviation of the metric from the reference metric, and $A^{ijkl}$ can depend on $\go$ but not on $g$.

Note that if there exists a rest configuration ($\go$ is an immersible metric), then we may choose the coordinates
$\vx$ to be the standard Cartesian coordinates on the undeformed configuration, thus setting $
\go_{ij}=\d_{ij}$. In such  case we may define the displacement vector $\vu=\vr-\vx$ to obtain
\[
\ep=\half (g-\mathbf{I})=\half((\vnabla \vr)^T\vnabla\vr-\mathbf{I})= \half(\vnabla \vu+(\vnabla
\vu)^T+(\vnabla\vu)^T\vnabla\vu),
\]
where $(\vnabla \vr)_{ij}=\pd r_i/\pd x_j$. We therefore identify $\ep$ as the \emph{Green-St. Venant strain
tensor}. The Frechet derivative of the energy density $w$ with respect to $\ep$ is the contravariant
\emph{second Piola-Kirchhoff stress tensor} \cite{Cia05}
\begin{equation}
S^{ij}=\frac{d w}{d \ep_{ij}}. \label{eq:S}
\end{equation}

For small strains we only need to determine the rank-four contravariant elasticity tensor $A^{ijkl}$.
Regardless of what $\go$ is at any given  point $p\in\o$, we may always choose a re-parametrization $\vx'$ such
that the reference metric with respect to the new (local) system of coordinates satisfies $\go'_{ij}=\d_{ij}$ at
$p$. If the medium is isotropic, then the tensor $(A')^{ijkl}$ at $p$ is isotropic in the Cartesian coordinates
$\vx'$, hence must be of the form
\begin{equation}
(A')^{ijkl}=\lambda \d^{ij}\d^{kl}+\mu(\d^{ik}\d^{jl}+\d^{il}\d^{jk}) \label{eq:A'}
\end{equation}
for some constants $\lambda$ and $\mu$ \cite{Cia05}. For a body with a reference rest configuration, we may
identify these constants as the Lam\'{e} coefficients.

It remains  to transform the contravariant tensor $A'$, defined on the local Euclidean coordinates $\vx'$, back
to the original curvilinear coordinates $\vx$ using the transformation rules for tensors,
\begin{equation}
A^{nmpq}=(\L^{-1})_{i}^n(\L^{-1})_{j}^m(\L^{-1})_{k}^p(\L^{-1})_{l}^q (A')^{ijkl},
\label{eq:A}
\end{equation}
where $\L=d\vx'/d\vx$ is the Jacobian of the transformation (see Appendix~\ref{tensors}). As the strain tensor
transforms with the jacobian
\[
g_{ij} -\go_{ij}=2\ep_{ij}=2\L^k_i\L^l_j\ep'_{kl}= \L^k_i\L^l_j(g'_{kl}-\d_{kl})= g_{ij}-\L^k_i\L^l_j\d_{kl},
\]
we obtain that $\L^k_i\L^l_j\d_{kl}=\go_{ij}$. Since all the orientation-preserving Cartesian coordinate
transformations differ only by a proper orthogonal rotation, this equation holds independently of the particular
local Cartesian set $\vx'$. The only implication of this calculation is that $\go$  must be symmetric and
positive-definite, i.e. it is indeed a metric. Yet, this metric is not required to be immersible in $\R^3$,
which is why we refer to our theory as ``incompatible" elasticity.

If we now define the reciprocal reference metric
by $\go^{jk}\go_{ki}=\d^j_i$, and substitute \eqref{eq:A'} in \eqref{eq:A}, using the fact that $(\L^{-1})^{ik}
(\L^{-1})^{jk}\delta_{kl} = \go^{ij}$, we obtain expression \eqref{A} for the energy density. As described in
Appendix \ref{tensors}, differentiation and the lowering and raising of indices are both defined with respect to
the reference metric. It should be emphasized that $\L_i^j$ and $\d_{ij}$ are not tensors in the sense defined in
Appendix~\ref{tensors} ($\d_{ij}$  is Kronecker's delta and not the lowered-index unit tensor). Moreover, given
a metric $g_{ij}$ there exists a reciprocal metric tensor $(g^{-1})^{ij}$ which is a contravariant tensor of
rank two and satisfies $(g^{-1})^{ij}g_{jk}=\d^i_k$, however it is not obtained by raising the indices of
$g_{ij}$, i.e. $(g^{-1})^{ij}\ne \go^{ik}\go^{jl}g_{kl}=g^{ij}$. The reference metric is the \textit{only} tensor
for which the inverse is obtained by raising both indices.

The equations of elastic equilibrium are obtained from the energy functional by a variational principle. We express the
energy as a functional of the metric tensor, $g$, yet variations of  $g$  must take into account that its
components satisfy six  differential constraints, which are the vanishing of the Ricci curvature tensor.
Alternatively, we may vary the configuration $\vr$, in which case the induced variation in $g$ trivially
satisfies the six constraints. Thus,
\[
\begin{split}
\d E &= \int_{\D}  \deriv{w}{\ep_{ij}} \d \ep_{ij} \dA\,d\vx
= \half \int_{\D} S^{ij} \d g_{ij} \dA\,d\vx \\
&= \int_{\D} S^{ij} \partial_i \vr\cdot\partial_j \d\vr\, \dA\,d\vx.
\end{split}
\]
Integrating by parts, and using the fact that
\[
\pd_j\pd_k \vr=\G^i_{jk} \pd_i\vr ,
\]
where
\[
\G^i_{jk} = \half (g^{-1})^{il}(\pd_j g_{kl} + \pd_k g_{jl} - \pd_l g_{jk})
\]
are the Christoffel symbols associated with the configuration $\vr$, we obtain after straightforward algebra the
following boundary value problem,
\begin{equation}
\begin{gathered}
\conab_j S^{ij}+ (\G^i_{jk}-\GO^i_{jk})S^{jk}=0 \qquad \text{ in  $\D$}\\
S^{ij}n_j=0 \qquad \text{ on $\pd\D$},
\end{gathered}
\label{eq:EulerLagrange}
\end{equation}
where
\[
\GO^i_{jk} = \half \go^{il}(\pd_j \go_{kl} + \pd_k \go_{jl} - \pd_l \go_{jk})
\]
are the Christoffel symbols associated with the reference metric, $n_j$ is the unit normal (in $\R^3$) to $\pd \D$,
and
\[
\conab_j S^{ij} = \frac{1}{\dA}\pd_j(\dA S^{ij})+\GO^i_{jk}S^{jk}
\]
is the covariant derivative with respect to the reference metric (see Appendix \ref{tensors}). As the elastic body
is immersed in $\R^3$ the six independent components of the symmetric Ricci curvature tensor of the metric $g$
\begin{equation}
\label{eq:Ricci}
\begin{split}
R_{li} &= \frac{1}{2}(g^{-1})^{kj}\left( \pd_k\pd_ig_{lj}-\pd_k\pd_j g_{li}+\pd_j\pd_l g_{ki}-\pd_i\pd_l
g_{kj}\right)\\
&+ (g^{-1})^{kj}g_{pq}\left(\G^p_{lj}\G^q_{ki}-\G^p_{kj}\G^q_{li}\right)
\end{split}
\end{equation}
must all vanish. The three equations \eqref{eq:EulerLagrange} together with the six immersibility conditions for
$g$ \eqref{eq:Ricci}, form a set of nine equations, for the six unknowns in $g$. There are two possible ways to
resolve this seemingly over-determination. The first is by noticing that the six independent components of the
Ricci curvature tensor satisfy differential relations: their derivatives are related through the second Bianchi
identity.
The second way of resolving this issue is by identifying the immersion $\vr$ as the three unknown functions, in
which case the six equations in \eqref{eq:Ricci} are solvability conditions for the PDE
\eqref{eq:EulerLagrange}. However, as the equations in $\vr$ are of higher order we need to supply additional
conditions, namely set the position and the orientation of the body, in order to obtain a unique solution for
$\vr$.

Eq. \eqref{eq:EulerLagrange} is our fundamental model for three-dimensional elasticity. The only (yet
fundamental) difference with standard models of finite displacement elasticity is that the reference metric does
not necessarily have an immersion in $\R^3$.

\section{The elastic theory of non-Euclidean plates}\label{elasticity plates}

We define a plate as
an elastic medium for which there exists a curvilinear set of coordinates in which the \emph{reference}
metric takes the form
\begin{equation}
\label{plate} \go_{ij}= \mymat{\go_{11} & \go_{12} & 0 \\ \go_{21} & \go_{22} & 0 \\ 0 & 0 & 1},
\quad\text{ where}\quad
\pd_3\go_{ij}=0.
\end{equation}
A plate is called \emph{even} if the domain $\D\subset \R^3 $ of the curvilinear coordinates can be decomposed
into $\D=\calS\times[-\frac{h}{2},\frac{h}{2}]$, where $\calS\subset \mathbb{R}^2 $ and $h$ is
constant. Thus an even plate is fully characterized by the metric of its mid-surface $x^3=0$. Let
\[
dA=\sqrt{\go_{11}\go_{22}-\left(\go_{12}\right)^2}dx^1dx^2
\]
denote an area element on the mid-surface, and $A=\int_\calS dA$ be the total area of the mid-surface. An
even plate will be called \emph{thin} if $h\ll\sqrt{A}$. A plate will be called \emph{non-Euclidean} if the
Ricci curvature tensor of its reference metric does not vanish. An equivalent condition is that the mid-surface
(considered as a two-dimensional manifold) has a non vanishing Gaussian curvature. A non-Euclidean plate has no
immersion with zero strain in $\R^3 $, i.e. the equilibrium state of a non-Euclidean plate must be a frustrated
state exhibiting residual stress. This statement is rather intuitive: If the plate fully complies with its given
two-dimensional metric, then it must assume a three-dimensional form that violates the invariance along the thin
direction. If, on the other hand, it remains planar, then it cannot comply with a non-vanishing Gaussian
curvature, hence it must contain in-plane deformations.

\subsection{The reduced energy density}

Although thin plates are three-dimensional bodies, one would like to take advantage of their large aspect ratio
and model them as two-dimensional surfaces, thus reducing the dimensionality of the problem. Ideally, one would
hope to obtain a reduced two-dimensional theory as an assumption-free small-$h$ limit of the three-dimensional
theory. Unfortunately, such an analysis is still lacking, and one must  introduce additional assumptions. We
adopt  the Kirchhoff-Love assumptions regarding the structure of the configuration metric $g$. The standard
formulation of the Kirchhoff-Love  assumptions is:

\begin{enumerate}
\item The body is in a state of plane-stress (the stress is parallel to the deformed mid-surface). \item Points
which are located in the undeformed configuration on the normal to the mid-surface at a point $p$, remain in the
deformed state on the normal to the mid-surface at $p$, and their distance to $p$ remains unchanged.
\end{enumerate}

The first assumption may be reformulated as
\[
S^{i3} = 0.
\]
In our case, where no reference configuration exists, the second assumption may be rewritten as
\[
g_{ij}=\mymat{ g_{\a\b}  & 0 \\  0 & 1},\quad \text{or equivalently} \quad \ep_{i3}=0,
\]
where following \cite{Cia05,Koi66} Greek indices $\alpha,\beta,\dots$ assume the values $\{1,2\}$. It is
important to note that the assumptions $S^{i3}=0$ and $\ep_{i3}=0$ represent two different elastic
problems---plane-stress versus plane-strain problems respectively. The two stand in contradiction  for all
$\lambda\ne 0$. As a result, the two assumptions do not ``commute", i.e. the order in which the two assumptions
are applied  is crucial. The key assumption is the first one, $S^{i3}=0$. It states that most of the elastic
energy is stored in lateral (in-plane) deformations of the various constant-$x^3$ planes. Estimates of
deviations from this assumption may be found in \cite{john}. Let $k_1$ and $k_2$ be the principal curvatures of
the mid-surface, $k_{max}=\max(k_1,k_2)$, and let $L$ be the smallest lateral length scale appearing in the
elastic equilibrium. It may be shown that the plane-stress approximation holds for
\[
k_{max} h \ll 1, \quad \text{and} \quad h \ll L.
\]
The second assumption, $\ep_{i3}=0$, is introduced only after we already have a reduced energy density,
containing only plane-stress contributions. It determines the actual three-dimensional configuration the body
assumes and the variation of the plane-stress along the thin dimension. It enables us to relate the elastic
energy density to geometric properties of the midplane which is considered as a two-dimensional surface.
Following \cite{Koi66} we denote by $\gamma$ the maximal plane-stress of the midplane and note that adding terms
of orders $\gamma^2$, $h k_{max} \gamma$ and $h^2 k_{max}^2$ to the energy density would not modify  the order of the approximation. Thus the second assumption may be considered as a subsidiary assumption, used to bring the
elastic energy density to the simplest consistent form. Although the assumptions are physically plausible,
reducing the three-dimensional energy functional into a two-dimensional functional by means of $\G$-convergence
would set the current theory of firmer grounds.

We now exploit the modified Kirchhoff-Love assumptions to derive a reduced two-dimensional model. Combining
\eqref{eq:S} and \eqref{A} and using the tensorial rules for raising indices we get
\[
S^{ij} = \lambda\go^{ij} \go^{kl} \ep_{kl} + 2\mu \go^{ik} \go^{jl} \ep_{kl} = \lambda \go^{ij} \ep^k_k + 2\mu
\ep^{ij}.
\]
From the first assumption, $S^{33} = 0$, and the fact that $\ep_k^k = \ep_\a^\a + \ep_3^3$ and $\ep_{33} = \ep_3^3 = \ep^{33}$, follows that
\begin{equation}
\ep_{33}=-\frac{\lambda}{\lambda+2\mu}\ep^\a_{\a}. \label{eq:ep33}
\end{equation}
We use \eqref{eq:ep33} to  rewrite the energy density \eqref{A} only in terms of the two-dimensional strain,
\[
w= \half A^{ijkl}\ep_{ij}\ep_{kl} = \half\brk{\lambda \ep^i_i\ep^k_k+2\mu \ep^k_j\ep^j_k}=
\mu\brk{\frac{\lambda}{\lambda+2\mu}\ep^\a_\a\ep^\b_\b+\ep^\a_\b\ep^\b_\a},
\]
or equivalently
\[
w= \half\A^{\a\b\g\d}\ep_{\a\b}\ep_{\g\d}, \qquad \A^{\a\b\g\d}=
2\mu\left(\frac{\lambda}{\lambda+2\mu}\go^{\a\b}\go^{\g\d}+\go^{\a\g}\go^{\b\d}\right).
\]
Note that as we contract the tensors $A$ and $\A$ with symmetric tensors we only retain their symmetric part. So
far we have only used the first of the Kirchhoff-Love assumptions.

We now use the second assumption to express the energy functional as a two-dimensional integral over the
mid-surface, by integrating $w$ over the thin coordinate $x^3$. As $g_{33}=\pd_3\vr\cdot\pd_3\vr=1$ and $g_{\a
3}=\pd_\a\vr\cdot\pd_3\vr=0$, we identify $\pd_3\vr = \hatn$ as the unit vector normal to the constant-$x^3$
surfaces. Moreover, it can be shown that $\pd_3\pd_3\vr=0$, implying that $\hatn = \hatn(x^1,x^2)$ is the unit
normal to the mid-surface, and $\pd_3\pd_3\pd_3 g_{\a\b}=0$.

The most general form of the metric is therefore given by
\begin{equation}
\label{metricZdependence} g_{\a\b}=a_{\a\b}(x^1,x^2)-2 x^3 \,b_{\a\b}(x^1,x^2) + (x^3)^2 c_{\a\b}(x^1,x^2).
\end{equation}
The tensors $a,b,c$ can be identified as follows: we define the mid-surface
\[
\vR(x^1,x^2) = \vr(x^1,x^2,0),
\]
and note that
\[
\pd_3 g_{\a\b}|_{x^3=0}
=\left[\pd_3\pd_\a\vr\cdot\pd_\b\vr+\pd_\a\vr\cdot\pd_3\pd_\b\vr\right]_{x^3=0}=-2\pd_\a\pd_\b\vR\cdot\hatn
\]
and
\[
\left.\pd_3\pd_3 g_{\a\b}\right|_{x^3=0}=2\pd_\a\hatn\cdot\pd_\b\hatn,
\]
which shows that $a,b,c$ are the first, second and third fundamental forms of the mid-surface. i.e.
\begin{equation}
a_{\a\b} = \pd_\a\vR\cdot\pd_\b\vR \qquad b_{\a\b}=\pd_\a\pd_\b\vR\cdot\hatn \qquad
c_{\a\b}=\pd_\a\hatn\cdot\pd_\b\hatn=(a^{-1})^{\g\d}b_{\a\g}b_{\b\d}. \label{eq:abc}
\end{equation}
A metric of the form \eqref{metricZdependence} with $a,b,c$ given by \eqref{eq:abc} corresponds to a
three-dimensional configuration of the form
\begin{equation}
\label{current-x3} \vr (x^1,x^2,x^3)=\vR(x^1,x^2)+x^3 \hatn(x^1,x^2).
\end{equation}

Having deduced the $x^3$ dependence of the metric in \eqref{metricZdependence},we may integrate the energy
density over the thin dimension,
\[
w_{2D}= \half \int^{\frac{h}{2}}_{-\frac{h}{2}}\A^{\a\b\g\d} \ep_{\a\b}\ep_{\g\d}dx^3
\]
which reduces to
\[
w_{2D}=\frac{h}{2} \A^{\a\b\g\d}\ep_{\a\b}^{2D}\ep_{\g\d}^{2D}
+\frac{h^3}{24}\A^{\a\b\g\d}\left(b_{\a\b}b_{\g\d}+\ep^{2D}_{\a\b}(a^{-1})^{\mu\nu}b_{\g\mu}b_{\d\nu}
\right)+\mathcal{O}(h^5),
\]
where $ \ep_{\a\b}^{2D}=\half(a_{\a\b}-\go_{\a\b})$ is the strain evaluated at the mid-surface. Omitting terms
of order five and higher in the thickness $h$, and neglecting $\ep$ with respect to the unit tensor yields the
final form of the reduced two-dimensional energy density,
\begin{equation}
w_{2D}=\frac{h}{2} \A^{\a\b\g\d}\ep_{\a\b}^{2D}\ep_{\g\d}^{2D} +\frac{h^3}{24}\A^{\a\b\g\d}b_{\a\b}b_{\g\d},
\label{eq:w2D}
\end{equation}
where
\[
\A^{\a\b\g\d}=\frac{Y}{1+\nu}\brk{\frac{\nu}{1-\nu}\go^{\a\b}\go^{\g\d}+\go^{\a\g}\go^{\b\d}}.
\]
We have introduced here  the physical constants $Y$ (Young's modulus) and $\nu$ (the Poisson ratio), defined by
\[
2\mu=\frac{Y}{1+\nu} \qquad\text{ and }\qquad \frac{\lambda}{2\mu+\lambda}=\frac{\nu}{1-\nu}.
\]
The total elastic energy is obtained by integration over the mid-surface
\begin{equation}
\label{EnergyD} E= \int_\calS w_{2D}\dA \,dx^1dx^2.
\end{equation}

We identify the two terms in \eqref{eq:w2D} as stretching and bending terms, respectively, and write the total
energy as
\[
E = h E_S + h^3 E_B,
\]
where
\[
E_S = \int_{\calS} w_S\, \dA\,dx^1 dx^2 \qquad E_B = \int_{\calS} w_B\, \dA\,dx^1 dx^2,
\]
and
\[
\begin{aligned}
w_S &= \frac{Y}{8(1+\nu)}\brk{\frac{\nu}{1-\nu}\go^{\a\b}\go^{\g\d}+\go^{\a\g}\go^{\b\d}}
(a_{\a\b}-\go_{\a\b}) (a_{\g\d}-\go_{\g\d}) \\
w_B &= \frac{Y}{24(1+\nu)}\brk{\frac{\nu}{1-\nu}\go^{\a\b}\go^{\g\d}+\go^{\a\g}\go^{\b\d}} b_{\a\b} b_{\g\d}.
\end{aligned}
\]
{\em Comments:}\\
\\
1. The quantities $E_S$ and $E_B$ are called the stretching and bending \emph{contents} (measures for the amount
of stretching and bending that do not vanish in the limit $h\to0$), and $w_S$ and $w_B$ are their respective
densities. By application of the Cayley-Hamilton theorem, the density of the bending content  can be rewritten
in the form
\[
w_B = \frac{Y}{24(1+\nu)}\brk{\frac{1}{1-\nu}(\go^{\a\b}b_{\a\b})^2 - 2\frac{|b|}{|\go|}}.
\]
2. A two-dimensional configuration has zero stretching energy if and only if $a_{\a\b} = \go_{\a\b}$, i.e., if
the two-dimensional metric coincides with the reference metric (such a configuration is an \emph{isometric
immersion} of $\go$).  In this case $(a^{-1})^{\a\b} = \go^{\a\b}$ and we identify the density of the bending
content as the density of the \emph{Willmore functional} \cite{willmore}
\begin{equation}
w_W = \frac{Y}{24(1+\nu)}\brk{\frac{4H^2}{1-\nu} - 2K},
\label{eq:willmore}
\end{equation}
where $K$ and $H$ are the Gaussian and mean curvatures of the mid-surface.

3. The total energy \eqref{EnergyD} is a functional of the mid-surface immersion $\vR$, i.e., $E = E(\vR)$. It
has two terms: the stretching energy, which scale linearly with $h$, and the bending energy, which scales like
the third power of $h$. The equilibrium configuration $\vR^*$ is the one that minimizes the energy functional.
For thin plates, the total  energy is dominated by the stretching term, and we expect the equilibrium
configuration to have a two-dimensional metric very close to the reference metric $\go$. For thick plates, it is
the bending energy which is dominant, and equilibrium is expected to have a minimal amount of bending.

\subsection{The reduced equilibrium equations}

As in the three-dimensional case, we can derive the Euler-Lagrange equilibrium equations that correspond to the
reduced energy functional \eqref{EnergyD} in two alternative ways. The first uses independent variations of the
six components of the symmetric tensors $a_{\a\b}$ and $b_{\a\b}$, adding three Lagrange multipliers to impose
the  three Gauss-Mainardi-Peterson-Codazzi (GMPC) equations:
\begin{equation}
\begin{gathered}
K=\frac{|b|}{|a|} =\frac{1}{2}(a^{-1})^{\a\b}\left(\pd_\g\G^\g_{\a\b}-\pd_\b\G^\g_{\a\g}
+\G^\g_{\g\d}\G^\d_{\a\b}-\G^\g_{\b\d}\G^\d_{\a\g}\right),\\
\pd_2 b_{\a 1}+\G^\b_{\a 1}b_{\b 2}=\pd_1 b_{\a 2}+\G^\b_{\a 2}b_{\b 1}.
\end{gathered}
\label{gmpc}
\end{equation}

The  GMPC equations are the necessary and sufficient condition for $a_{\a\b}$ and $b_{\a\b}$ to be the first and
second fundamental forms of a surface in $\R^3$. It is noteworthy that the satisfaction of the GMPC equations is
a sufficient condition for the immersibility of a metric of the form \eqref{metricZdependence} \cite{Cia05}.
Again this mathematical result is rather intuitive: If the tensors $a_{\a\b}$ and $b_{\a\b}$ satisfy the GMPC
equations, then there exists a mid-surface $\vR(x^1,x^2)$, for which they constitute the first two fundamental
forms. If such a surface exists then the explicit construction \eqref{current-x3} ensures the existence of an
immersion in $\R^3$ of the three-dimensional body.

The second and more natural path is to preform variations in the mid-surface $\vR$, \cite{Cia05,Koi66}. Let us
define the reduced two-dimensional stress and moment tensors by
\[
s^{\a\b}=\frac{\pd w_{2D}}{\pd \ep^{2D}_{\a\b}}=h \A^{\a\b\g\d}\ep^{2D}_{\a\b} \qquad m^{\g\d}=\frac{\pd
w_{2D}}{\pd b_{\g\d}}=\frac{h^3}{12}\A^{\a\b\g\d}b_{\a\b}.
\]
Consider then a variation $\vR\rightarrow\vR+\d\vR$. To first order in $\d\vR$ we have
\[
\begin{aligned}
\d \ep^{2D}_{\a\b}&=\half\brk{\pd_\a\vR\cdot\pd_\b\d \vR+\pd_\b\vR\cdot\pd_\a\d \vR} \\
\d b_{\a\b}&=\pd_\a\pd_\b \d\vR \cdot \hatn+\pd_\a\pd_\b \vR \cdot \d\hatn=\pd_\a\pd_\b \d\vR \cdot
\hatn-\G^\g_{\a\b}\hatn \cdot \pd_\g \d \vR,
\end{aligned}
\]
where from now on the Christoffel symbols $\G^{\g}_{\a\b}$ are defined with respect to the two-dimensional
surface $\vR$ (they are the restriction of $\G^i_{jk}$ to the indices $\{1,2\}$). The resulting
variation in the energy is
\[
\d E= \int_\calS \brk{s^{\a\b}\d\ep^{2D}_{\a\b}+m^{\a\b}\d b_{\a\b}}\sqrt{\go}\,dx^1dx^2.
\]
Integrating by parts gives the following equation,
\begin{equation}
\label{equaeq}
\begin{split}
0 &= \conab_\a\left(\conab_\b m^{\a\b}+(\G^\a_{\d\b}-\GO^\a_{\d\b})m^{\d\b}\right)-s^{\a\b}b_{\a\b}-m^{\a\b}c_{\a\b} \\
0 &= \conab_\b\left(s^{\a\b}+m^{\mu\b}(a^{-1})^{\g\a}b_{\mu\g}\right)+(\G^\a_{\d\b}-\GO^\a_{\d\b})
\left(s^{\d\b}+m^{\mu\b}(a^{-1})^{\g\d}b_{\mu\g}\right) \\
&+\left(\conab_\b m^{\mu\b}+(\G^\mu_{\d\b}-\GO^\mu_{\d\b})m^{\d\b}\right)(a^{-1})^{\g\a}b_{\g\mu},
\end{split}
\end{equation}
and boundary conditions:
\[
\begin{aligned}
0 &= n_\a n_\b m^{\a\b} \\
0 &= n_\b\left(s^{\d\b}+(a^{-1})^{\mu\d}b_{\mu\a}m^{\a\b}\right) \\
0 &= n_b\left(\conab_\a m^{\a\b}+(\G^\b_{\a\d}-\GO^\b_{\a\d})m^{\a\d}\right),
\end{aligned}
\]
where
\[
\begin{aligned}
\conab_\b V^\b &= \frac{1}{\dA}\pd_\b(\dA V^\b) \\
\conab_\b M^{\a\b} &= \frac{1}{\dA}\pd_\b(\dA M^{\a\b})+\GO^\a_{\b\d}M^{\b\d}.
\end{aligned}
\]
The three equations \eqref{equaeq} (in the second equation $\a=1,2$ is a free index), supplemented by the three
GMPC equation \eqref{gmpc}, form a boundary value problem for $a_{\a\b}$ and $b_{\a\b}$ as well as an
integrability condition for $\vR$.

\section{Example: A spherical plate annulus}

\subsection{Axially symmetric case}

The reduced two-dimensional equilibrium equations \eqref{equaeq} are highly nonlinear equations in the six
variables $s^{\a\b}$, $m^{\a\b}$.  A tractable set of equations may be obtained if, for example, symmetries are
imposed. Let us set $x^1=r, \, x^2= \theta$ (polar coordinates) and consider a reference metric of the following
form:
\begin{equation}\label{equ:semigeodesic}
\go_{\a\b}(r,\theta)= \mymat{1 & 0 \\0 & \Phi^2(r) }
\end{equation}
In this case, the Gaussian curvature of the mid-surface is $K = -\Phi_{rr}/\Phi$, where we now use subscripts
to denote differentiation. Recall that the corresponding three-dimensional reference metric $\go_{ij}$ given by
\eqref{plate} can be immersed in $\R^3$ only if $K=0$.

We seek solutions in the form of a body of revolution
\[
\vR(r,\theta)=(\phi(r)\cos\theta,\phi(r)\sin\theta,\psi(r)).
\]
For such configurations the GMPC equations are satisfied trivially. The first and second fundamental forms are
given by
\[
a_{\a\b}= \mymat{ \phi_r^2+\psi_r^2 & 0 \\ 0& \phi^2 } \quad\text{ and }\quad
b_{\a\b}=\frac{1}{\sqrt{\phi_r^2+\psi_r^2}}\mymat{ \psi_{rr}\phi_r-\phi_{rr}\psi_r &0\\ 0& \phi\psi_r }.
\]
If we define $\psi_r=\phi_r \zeta$ (which implies that $\psi_{rr}\phi_r-\psi_r\phi_{rr}=\phi_r^2 \zeta_r$),
then, substituting the fundamental forms into the two-dimensional energy density \eqref{eq:w2D}, we obtain the
following expression for the energy,
\begin{equation}\label{eq:Eaxial}
E = \frac{\pi Y}{4(1-\nu^2)} \int_{\calS} w_{2D} \Phi\,dr,
\end{equation}
where
\[
w_{2D} = h w_S + h^3 w_B,
\]
and
\[
\begin{aligned}
w_S &=
2\nu\left(\phi_r^2(1+\zeta^2)-1\right)(\phi^2/\Phi^2-1) + \left((\phi_r^2(1+\zeta^2)-1)^2+(\phi^2/\Phi^2-1)^2\right) \\
w_B &= \frac{2\nu}{3}\frac{1}{(1+\zeta^2)} (\phi\phi_r \zeta \zeta_r/\Phi^2) +
\frac{1}{3}\frac{1}{(1+\zeta^2)}\left( (\phi_r \zeta_r)^2+(\phi \zeta/\Phi^2)^2\right)
\end{aligned}
\]
are the densities of the stretching and bending contents. Note that the introduction of $\zeta$ yields an energy
density that only includes first-derivatives of $\phi$, and $\zeta$.

The minimum energy configuration balances the contributions from both stretching and bending terms. Upper bounds
on the minimum energy can be derived by considering the two extreme cases, which contain no stretching and no
bending, respectively. Consider first stretch-free configurations, $w_S=0$, which occur when the two-dimensional
metric $a_{\a\b}$ coincides with the two-dimensional reference metric, $\go_{\a\b}$, i.e., when
\[
\phi = \Phi \qquad\text{ and }\qquad \phi_r^2  + \psi_r^2 = \phi_r^2(1+\zeta^2) = 1.
\]
Thus, there exists a unique axially symmetric isometric immersion
(however, infinitely many non-axisymmetric isometric immersions may exist).
The density of the bending content of this
isometry reduces to
\[
w_B = - \frac{2\nu}{3} \frac{\Phi_{rr}}{\Phi} + \frac{1}{3}  \brk{ \frac{ \Phi_{rr}^2}{1 - \Phi_r^2} + \frac{1 -
\Phi_r^2}{\Phi^2}},
\]
which is the density $w_W$ of the Willmore functional. Integration of this density provides a first upper bound
on the equilibrium energy.

Consider next bending-free configurations, $w_B=0$, obtained if and only if $\zeta=0$. This implies that
$\psi_r=0$, i.e., a flat radially symmetric surface. The density of the stretching content reduces to
\[
w_S = 2\nu(\phi_r^2-1)(\phi^2/\Phi^2-1) + (\phi_r^2-1)^2+(\phi^2/\Phi^2-1)^2.
\]
Note that there are infinitely many axially symmetric configurations for which the bending content vanishes.
Finding the configurations that minimizes the stretching energy is equivalent to solving the axially symmetric
plane-stress problem, which can be achieved numerically.

\subsection{Numerical results}\label{numerics}

As an example, we consider the case where the two-dimensional reference metric $\go_{\a\b}$ is that of a sphere,
$\Phi(r) = \sin r$, and the domain is an annulus,
\[
r\in [r_{\text{min}},r_{\text{max}}] \subset (0,\pi/2).
\]
The stretch-free configuration is a punctured spherical cap and its experimental realizations are shown in
Figure \ref{fig:caps}.

The minimizer of the energy functional \eqref{eq:Eaxial} was computed numerically for the parameters $\nu=0.5$,
$r_{\text{min}}=0.1$ and $r_{\text{max}}=1.1$. The elastic modulus $Y$, which is immaterial to the equilibrium
shape, was set such that the pre-factor $\pi Y/4(1-\nu^2)$ equals one. As expected, for values of $h$ above the
buckling transition ($h_B\approx 0.3$) the solution is that of a flat plate, whereas for values of $h$ under the
buckling transition, the plate is close to spherical.

\begin{figure}[h]
\putfig{2.7}{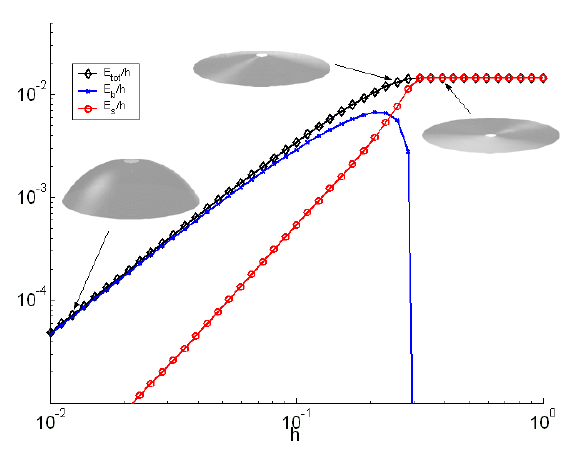} \caption{Energy scaling in positively curved discs. Total elastic energy,
stretching energy and bending energy versus the thickness $h$, for  non-Euclidean discs with the reference metric
and lateral dimensions that are described in the text. All three energies are divided by $h$. The three rendered
configurations correspond to (from right to left): a flat configuration, a weakly buckled configuration, just
below the buckling threshold, and a fully buckled, almost isometric configuration (color online).}
\label{fig:energy}
\end{figure}

In Figure~\ref{fig:energy} we plot the stretching energy (red circles), the bending energy (blue crosses) and
the total energy (black diamonds) versus the plate thickness $h$; all three energies were scaled by $1/h$.
Except for a narrow transition region near the buckling threshold, the total energy is dominated by either the
stretching energy or the bending energy. As one would expect, the bending energy drops to zero above the
buckling threshold (large thickness). However, below the buckling threshold, as $h\to0$, the stretching energy
drops to zero much more rapidly than the bending energy. This last observation is in fact surprising, as
naively, one would expect equilibrium to be attained when both stretching and bending energy are ``equally
partitioned" \cite{Ven04}.

\begin{figure}[h]
\putfig{2}{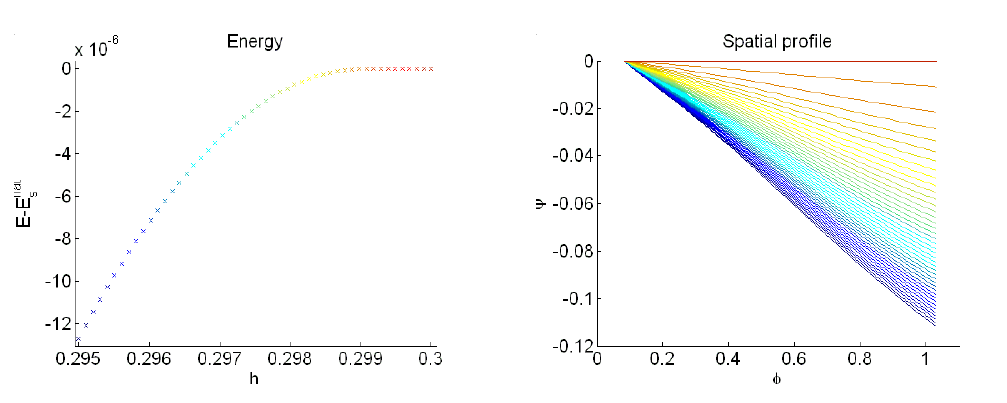} \caption{The Buckling transition. The energy variation divided by the thickness,
$(E-E_s^{flat)})/h$ (left), and spatial profile (right) near the buckling threshold as a function of the
thickness. Every profile in the right figure corresponds to an energy variation bearing the same color in the
left figure. All configurations are flatter than the center disc in Figure~\ref{fig:energy} (note the difference
in scales of the axis on the right).}
\label{fig:buckling}
\end{figure}

In Figure \ref{fig:buckling} the spatial profile (a cross-section) of the elastic equilibrium configuration is
shown. The transition from flat to buckled configurations occurs continuously, hence
the buckled states, close to the buckling threshold, are nearly
planar. This supports the validity of theories that assume small deflections from a plane (such as the
FVK model) for predicting the buckling threshold. As the thickness is further reduced, the plate approaches the
stress-free (isometric) configuration very fast. The assumption of small deflections from a plane fails for such
configurations.

The minimal bending content, $E_B^0$, of the stretch-free configuration, and the minimal stretching content, $E_S^\infty$, of the zero bending configuration yield a crossover length scale: $h_C=\sqrt{E_S^\infty/E_B^0}$. Linear analysis about a flat surface gives another length scale, the buckling threshold thickness $h_B$. We expect the scenario
depicted in Figure \ref{fig:energy} to be valid for bodies in which these two length scales are relatively
close. However, there are reference metrics (specifically, hyperbolic), for which all isometric immersions are
convoluted, i.e. $E_B^0$ is very large. For such bodies one may obtain $h_C\ll h_B$. When this occurs, the
transition region may expand. For such bodies the scaling of the elastic equilibrium energy with the thickness
will be very different from the one appearing in Figure \ref{fig:energy}.

\section{Conclusion}

Natural growth of tissue as well as the plastic deformation of solids are examples of local shaping mechanisms of
elastic bodies. In general, the local nature of such growth processes excludes the existence of stress-free
configurations.
This is the main reason why current elastic theories cannot handle properly
such shaping mechanisms. In this work we derived a reduced 2D model for a class
of thin plates with residual stresses, which we named ``non-Euclidean plates''. Such plates are uniform across
their thin dimension, but their 2D geometry is non-Euclidean. Their complicated 3D configurations cannot be obtained
from existing 2D models of elasticity. Our derivation is based on a covariant formulation of 3D linear
elasticity. It does not require the existence of a reference stress-free configuration, but only a
3D ``reference metric'' tensor, which is determined by the growth. We use this formalism together with
the Kirchhoff-Love assumptions to derive a 2D energy functional. Like preceding theories, this functional
decouples into bending and stretching terms. The bending term scales like the third power of the thickness and
depends on surface curvature. The  stretching term scales linearly with the thickness and increases with
in-plane strain, which is nothing but the difference between the 2D metric tensor of a configuration and the 2D
reference metric. Our theory is valid for large rotations and displacements and arbitrary intrinsic metrics.

The numerical results presented in Figure \ref{fig:energy} suggest that in the general case there is no
equipartition between bending and stretching energies. This in turn supports the treatment of very thin bodies
as inextensible. Not only the equilibrium three-dimensional configuration is dominated by the minimization of
the ``small" bending energy term, but the total elastic energy is dominated by it too. The estimate of what
thickness should be considered as thin involves the introduction of a new length scale $h_C$, which is smaller
than the buckling threshold thickness. The square of this new length scale, $h_C^2$, is inversely proportional
to the minimum of the Willmore functional for the prescribed $2D$ geometry. This length scale differentiates
between two types of surface geometries. Surfaces which may be isometrically immersed with a moderate bending
content, for which $h_C$ is close to the buckling threshold thickness, will follow the shaping scenario and
energy profile described in Figures \ref{fig:energy} and \ref{fig:buckling}. Surfaces for which all isometric
immersions have high bending contents (as is the case for some hyperbolic surfaces) may exhibit very different
shaping scenarios and energetic landscapes.

The theory can be further elaborated and generalized to describe a wider range of growing bodies. We believe,
however, that already in its current stage, it is a powerful tool for studying the growth of leaves and other
natural slender bodies.

{\bfseries Acknowledgments}
This work was supported by the United States-Israel Binational Foundation (grant no. 2004037)
and the MechPlant project of European Unions New and Emerging Science and Technology program.
RK is grateful to M.R. Pakzad and M. Walecka for pointing out an error in the original manuscript.

\appendix
\section{Tensors, vectors, scalars and the covariant derivative}
\label{tensors}

As our treatment of elastic bodies involves the simultaneous use of two different metrics, we find it important
to provide a brief summary of differential geometry in the context of the current work. In the following
treatment we do not consider the most general setting but only three-dimensional manifolds immersed in $\R^3$.

Let the immersed manifold $\o \subset \R^3$ be the current configuration of an elastic body. A global
parametrization of $\o$ is a one-to-one map $\vr :\D \rightarrow\o $ from a domain $\D\subset\R^3$. Let $\vr ':
\D ' \rightarrow \o $ be a different global parametrization of the current configuration. The composition
$\mathbf{h}= \vr'^{-1}\circ \vr:\D\rightarrow\D'$ is called a coordinate transformation. The coordinate
transformation gradient, often denoted by $\L^j_i=\pd x'^j/\pd x^i $, is simply the Jacobian matrix of the
transformation $\mathbf{h}$, i.e. $\L=\pd \mathbf{h}/\pd \mathbf{x}$. The inverse transformation gradient is
$\left(\L^{-1}\right)^i_j=\pd x^i/\pd x'^j$.

A scalar is a function $\Phi:\o\to\R$. Given a parametrization $\vr:\D\to\o$, a scalar $\Phi$ induces a function
$\phi:\D\to\R$ defined by $\phi(\mathbf{x})=\Phi(\vr(\mathbf{x}))$. Given another parametrization
$\vr':\D'\to\o$ with the coordinate transformation $\mathbf{h}:\D\to\D'$, the relation between the induced
functions $\phi$ and $\phi'$ is $\phi'(\mathbf{x}')=\phi(\mathbf{h}^{-1}(\mathbf{x}'))$. By a slight abuse of
terminology we also call the functions $\phi$ and $\phi'$ scalars

A vector is a function $\V$ from the manifold $\o$ to the local tangent space of the manifold which in our case
is $\R^3$, $\V:\o\to\R^3$. Note that we cannot perform vector operations on pairs of vectors defined at two
different points in $\o$, as they belong to different tangent spaces (or equivalently different copies of
$\R^3$). Given a parametrization we may construct a basis $\e_i=\pd\vr/\pd x^i$ for each tangent space. With
respect to this basis we may write any vector as $\V=V^i\e_i$. The three functions $V^i:\D\to\R$ are called the
contravariant components of the vector $\V$. Again by an abuse of terminology the triplet $V^i,\, i={1,2,3}$ is
called a contravariant vector. It is easy to prove that under a coordinate transformation, a contravariant
vector transforms with the inverse transformation gradient, $V^i=\left(\L^{-1}\right)^i_j V'^j$, where the
left-hand side is estimated at a point $\mathbf{x}$ while the right-hand side is estimated at the corresponding
point $\mathbf{x}'=\mathbf{h}(\mathbf{x})$.

We next define the dual vector space, namely the space of covariant vectors. However, as there are many ways to
define an inner product on the tangent space, there are just as many ways to define the dual vector space. The
most natural inner product is the inner product induced from $\R^3$. In such a case, we define a dual base $\e^j$
by the condition $\e^j\cdot\e_i=\d^j_i$, where $ \cdot $ is the Euclidean product in $\R^3$. Any vector in the
tangent space may now be decomposed with respect to this basis, $\V=V_i\e^i$. The triplet $V_i$ is called a
covariant vector. Under a coordinate transformation covariant vectors transform with the transformation gradient
$V_i=\L^j_i V'_j$. The inner product in the local tangent space induces an inner product on the space of
contravariant vectors and the mapping of contravariant vectors to their covariant duals by
\[
\V\cdot \mathbf{U}=V^iU^j\e_i\cdot\e_j=g_{ij}U^iV^j =V^iU_j\e_i\cdot\e^j=V^iU_j\d^j_i=U_jV^j,
\]
where $g_{ij}=\e_i\cdot\e_j$ is called the Euclidean metric of $\o$ with respect to the given coordinate system.
The tensor $g_{ij}$ transforms covariantly in both indices, i.e. $g_{ij}=\L^k_i \L^l_j g'_{kl}$. We have
identified each contravariant vector $V^i$ with a (covariant) vector from the dual space $V_i=g_{ij}V^j$, which
is called a covariant vector. The contraction of a covariant and a contravariant vector $V^iU_i$ yields a
scalar. We may choose other inner products on the space of contravariant vectors, leading to different
definitions of the dual space. Let $\go_{ij}$ be a positive definite symmetric tensor, which transforms under a
coordinate transformation by $\go_{ij}=\L^k_i \L^l_j \go'_{kl}$ (i.e. covariantly in both indices). The
operation $\langle,\rangle:\R^3\times\R^3\to\R$ given by $\langle\mathbf{U},\V\rangle=\go_{ij}U^iV^j$ defines an
inner product on the space of contravariant vectors. For every contravariant vector there corresponds a
covariant dual given by $V_j=\go_{ij}V^i$. The tensor $\go$ is called the covariant metric on $\o$.

Given a parameterized manifold $\vr:\D\to\o$ one may easily prove that the gradient of a scalar
$V_i=\pd_i\phi=\pd \phi/\pd x^i$ is a covariant vector. However in order to differentiate vectors we need to
compare vectors that belong to different tangent spaces. To do so we use parallel transport of one of the
vectors to the point where the other vector is defined. To give only a notion of what parallel transport is, we
say that it will be transporting the vector along a "straight line", keeping a constant angle between the line
and the vector. Both concepts, angles between a curve and a vector, as well as ``straight lines" (geodesics),
are defined by the covariant metric tensor. Thus, while the differentiation of a scalar is independent of the
metric, the differentiation of a vector depends on the  metric. It may be shown that the parallel transport
procedure results in the following definition of the covariant derivative.
\[
\conab_i V_j=\pd_i V_j-\GO^k_{ij}V_k,
\]
where
\[
\GO^i_{jk} = \half \go^{il}(\pd_j \go_{kl} + \pd_k \go_{jl} - \pd_l \go_{jk}).
\]
One may verify that $\conab_i V_j$ transforms covariantly in both indices under a coordinate transformation. The
covariant differentiation of a contravariant vector is given by
\[
\conab_i V^j=\pd_i V^j+\GO^j_{ik}V^k.
\]
Note that $\GO^i_{jk}$ is not covariant or contravariant in any of its components. Henceforth, we will use the
term tensors to refer to multidimensional arrays for which all indices transform covariantly or contravariantly,
thus $\GO$ is not a tensor. One may easily verify that the multiplication or contraction of tensors results in a
tensor. The differentiation of a tensor should be treated as if the tensor is an external product of vectors and
apply the covariant derivative through the Leibnitz product rule. For example in the two-dimensional case we
have
\[
\conab_k M_{ij}=\pd_k M_{ij}-\GO^l_{kj}M_{il}-\GO^l_{ki}M_{jl}.
\]

In general, when working with explicit parameterizations we need, in order to prove that a certain parameter is
a tensor (e.g. a scalar or a covariant vector),  to prescribe it for all possible parameterizations, and show
that it obeys the correct transformation rules. This is the case for the current metric
$g_{ij}=\pd_i\vr\cdot\pd_j\vr$. It is defined for all possible parameterizations and obeys the covariant
transformation rules. As the reference metric coincides with the current metric (for a local stress-free
configuration), we have that $\go$ is also a rank-two covariant tensor. However some quantities are tensorial by
definition, for example $S^{ij}=d w/d \ep_{ij}$, which is the derivative of a scalar with respect to a covariant
tensor. For such quantities we may determine their value for one (convenient) parametrization, and obtain their
value for all other parameterizations through the tensorial transformation rule. This is the case for the
elastic tensor $A^{ijkl}$, as may be observed in \eqref{eq:A}.

\bibliographystyle{unsrt}

\end{document}